\documentclass[showpacs,preprintnumbers,amsmath,amssymb,11pt]{revtex4}
\usepackage{graphicx}
\newcommand{\s}{\ensuremath{\psi(t,r)}}
\newcommand{\n}{\ensuremath{\nu(t,r)}}
\newcommand{\T}{\ensuremath{\theta}}
\newcommand{\pt}{\ensuremath{p_{\theta}}}
\newcommand{\pr}{\ensuremath{p_r}}
\newcommand{\e}{equation$\;$} 
\newcommand{\M}{\ensuremath{{\cal M}}}
\newcommand{\Ma}{\ensuremath{ m_{0}}}
\newcommand{\Mb}{\ensuremath{ m_{2}}}
\newcommand{\prz}{\ensuremath{p_{r_0}}}

\newcommand{\ptz}{\ensuremath{p_{\theta_0}}}
\newcommand{\ptt}{\ensuremath{p_{\theta_2}}}

\newcommand{\X}{\ensuremath{{\cal X}}}

\def\be{\begin{equation}}
\def\eq{\end{equation}}

\begin{document}
\preprint{}
\title{Rebounce and Black hole formation in a Gravitational Collapse Model
 with Vanishing Radial Pressure}
\author{Ashutosh Mahajan}
\email{ashutosh@tifr.res.in}
\author{Pankaj S. Joshi}
\email{psj@tifr.res.in}
\affiliation{Tata Institute of Fundamental Research\\
Homi Bhabha Road\\ Mumbai 400 005, India}

\begin{abstract}
We examine spherical gravitational collapse of a matter 
model with vanishing radial pressure and non-zero tangential pressure. 
It is seen analytically that the collapsing cloud either forms a 
black hole or disperses depending on values of the initial parameters 
which are initial density, tangential pressure and velocity profile of the 
cloud. A threshold of black hole formation is observed near which
a scaling relation is obtained for the mass of black hole, assuming 
initial profiles to be smooth. The similarities in the behaviour of 
this model at the onset of black hole formation with that of numerical 
critical behaviour in other collapse models are indicated.  
\end{abstract}
\maketitle

\section{Introduction}

Astrophysical black holes have a minimum mass, known as the 
Chandrashekhar mass. Realistic matter in a star is almost stationary 
at the initial epoch of its collapse and has a characteristic scale 
which depends on the properties of matter. If one does not restrict to 
stationary realistic matter, then in principle one should be able to 
produce arbitrary small mass black holes from initially collapsing cloud, 
as there will not be any characteristic scale present in the cloud 
and the positive pressure in the cloud has tendency to push the matter 
in radially outward direction in spherically symmetric situation.

There can be very many possible equation of states with pressure 
present that the collapsing matter can satisfy, but the 
non-linear Einstein equations are tractable analytically only for a
very few equations of state. Here, we examine a spherically symmetric 
collapse model, in which the radial pressure vanishes and non-zero 
tangential pressure is present in the cloud. 
The Einstein cluster is such an example with vanishing radial pressure, 
where a cloud of rotating particles whose motion is sustained by 
an angular momentum has an average effect of creating a non-zero 
tangential stress. Such a static system was first introduced by Einstein 
\cite{ein}, 
which was later generalized to non-static 
case 
\cite{Bondi}.  
Subsequently, the gravitational collapse models with a vanishing
radial pressure have been investigated extensively by Magli 
\cite{magli} 
and others  
\cite{tan} 
for the purpose of examining the validity or otherwise of the 
cosmic censorship conjectures and to understand the black hole and
naked singularity formation in gravitational collapse.

We observe an intriguing feature in this model that it has a threshold of black hole formation and it shows a power law behaviour for the mass of the black hole near this threshold. We work in the comoving coordinates $(t,r,\theta,\phi)$ and 
construct an effective potential for the collapsing shells, with the 
aid of which one can see transparently the various possible evolutions 
of the collapsing shells. We demonstrate for a set of initial data 
that the collapsing cloud either forms a black hole or completely disperses, 
depending on the values of initial parameters. We point out similarities in behavior 
of this model at the threshold of black hole formation with already 
established critical behavior seen in other matter models 
\cite{radiation}.

The outline of the paper is as follows. In Section II, we discuss 
the collapse equations and regularity conditions and in Section III, a 
tangential pressure model is constructed. In Section IV, various possible 
dynamical evolutions are illustrated and a scaling relation for the 
black hole mass is obtained. Discussion and conclusions are outlined in 
Section V.


\section{Einstein Equations, Regularity and Energy Conditions}

We consider four dimensional spherically symmetric metric in 
comoving coordinates,  
\begin{equation}
ds^2= -e^{2\n}dt^2 + e^{2\s}dr^2 + R^2(t,r)d\Omega^2,
\label{metric}
\end{equation}
where $d\Omega^2$ is the line element on two-sphere. The energy-momentum 
tensor is then diagonal for any general {\it Type I} collapsing matter field
and is given as $T^t_t=-\rho;\; T^r_r=p_r;\; T^\T_\T=T^\phi_\phi=p_\T$.
This is a general class of matter fields that includes
many known physical forms of matter
\cite{he}.
The quantities $\rho$, $p_r$ and $p_\T$ are the density, 
radial and tangential pressures respectively. We take the matter 
field to satisfy the {\it weak energy condition}, that is, the energy 
density  measured by any local observer be non-negative. Then for 
any timelike vector $V^i$ we have
$T_{ik}V^iV^k\ge0$, {\it i.e.} $\rho\ge0;\; \rho+p_r\ge0;\; 
\rho+p_\T\ge0$.
The dynamical evolution of the system is determined by the Einstein 
equations and for metric (\ref{metric}) these are given as
\be
\rho = \frac{F'}{R^2R'}, \; \; \; \;
  p_{r}=-\frac{\dot{F}}{R^2 \dot{R}},
\label{t6}
\eq
\be
\nu'(\rho+ p_{r})=2(p_{\theta}-p_{r})\frac{R'}{R}-p_{r}',
\label{t7}
\eq
\be
-2 \dot{R'}+R'\frac{\dot{G}}{G}+\dot{R}\frac{H'}{H}=0,
\label{t8}
\eq
\be
G-H=1 - \frac{F}{R},
\label{t9}
\eq
\\
where $(\,\dot{}\,)$ and $(')$ are partial derivatives 
with respect to $t$ and $r$ respectively and
\be
G(t,r)=e^{-2\psi}R'^2, \; \; H(t,r)=e^{-2\nu}\dot{R}^2.
\eq

The function $F(t,r)$ is twice the Misner-Sharp mass for the 
collapsing cloud, which gives the total mass within the shell 
of comoving radius $r$ at time $t$ 
\cite{misner}.
Regularity at the initial epoch implies $F(t_i,0)=0$, that is, the 
mass function vanishes at the center of the cloud. It is seen from 
($\ref{t6}$) that the density of matter blows up when $R=0$ or $R'=0$. Here 
$R'=0$ corresponds to a {\it shell-crossing} singularity.
We use the scaling independence of the coordinate $r$ to write,
$R(t,r)=rv(t,r)$ and we have
\begin{eqnarray}
v(t_i,r)=1; & v(t_s(r),r)=0; & \dot{v}(t_i,r)<0,
\label{v}
\end{eqnarray}
where $t_i$ and $t_s$ are the initial and  singular   
epochs respectively.  The condition $\dot{v}(t_i,r)<0$ 
signifies initially collapsing shells. At the initial epoch we
have $R=r$, {\i.e.} $v(t,r)=1$, and at the singularity $R=0$ and
$v=0$. At all other epochs $v$ has a non-zero finite value for 
all $r$ with $0<r<r_b$, where $r_b$ is the boundary of the cloud.

From the point of view of dynamic evolution of 
initial data from $t=t_i$, we now have five arbitrary 
functions, given by $\nu(t_i,r)=\nu_0(r), \psi(t_i,r)=\psi_0(r), 
\rho(t_i,r)=\rho_0(r), p_r(t_i,r)=p_{r_0}(r), \pt(t_i,r)=p_{\T_0}(r)$.
They are all not independent, (\ref{t7}) gives a relation
for them. To preserve the regularity and smoothness of  
initial data we assume that the gradients of pressures vanish at the 
center, $\prz'(0)=\ptz'(0)=0$. Thus we have a total of five field 
equations with seven unknowns $\rho$, $p_r$, $\pt$, $\psi$, $\nu$, 
$R$, and $F$, giving a freedom of choice of two free functions to 
complete the system. This choice, subject to the given initial data 
and weak energy condition, determines the matter distribution and 
metric of the space-time, leading to a particular dynamical
collapse evolution of the initial data.

\newpage

\section{Tangential Pressure Model}

In this framework, we now study a class of collapse models 
with vanishing radial pressure and non-vanishing tangential pressure,
which have been studied extensively in recent years 
\cite{magli,tan}. 
For the sake of 
transparency, we construct and consider below an explicit collapse 
solution with a non-vanishing tangential pressure. We choose the allowed 
two free functions, $F(t,r)$ and $\n$, in the following way. The vanishing 
or a constant $p_r$ implies that the mass function $F$ has to be 
necessarily of the 
form $F(t,r)=r^3\M (r)$, where $\M$ can be any general function 
of $r$, and we also choose $\n=\nu_0(R)$.

Consider $\M$ now be of the form
\be
\M(r)=\Ma e^{-\Mb r^2},
\label{mass1}
\eq
where $\Ma$ and $\Mb$ are positive constants.
Using the above form of mass function in equation (\ref{t6}), we get
$\pr=0$, {\it i.e.} the radial pressure vanishes identically, and
the density at initial epoch is given by
$\rho_0(t_{i},r)=(3-2\Mb r^2)\Ma e^{-\Mb \,r^2}$.
We also take the initial tangential pressure to be of the form
\be
\pt(t_{i},r)=e^{\,\ptt r^2}-1,
\label{ptheta0}
\eq
such that both $p_{\theta}$ and $p_{\theta}'$ vanish at the center.
We note that while the above profiles are chosen in order
to give one particular solution, the following analysis is valid for any 
smooth initial data with initial density and pressure expressed 
in even powers of $r$. In general, as $v\rightarrow 0$, 
$\rho\rightarrow\infty$ and the density blows up at $R=0$, which is 
a curvature singularity as expected.
Using $\nu=\nu_0(R)$ in \e (\ref{t8}), we get
\begin{equation}
G(t,r)=f(r)e^{2\nu_0(R)},
\label{G}
\end{equation}
where $f(r)$ is another arbitrary function of $r$. 
In analogy with dust collapse models, we write
\begin{equation}
f(r)=1+r^2b(r),
\label{veldist}
\end{equation}
where $b(r)$ is the energy distribution function for the 
collapsing shells. We take it to be a smooth function, 
$b(r)=-b_{0}+b_{2}r^2+\cdots$. Using $\nu$ in (\ref{t7}), 
the equation of state turns out to be $2\pt = R \, \nu_{,R} \, \rho$.
Finally, using (\ref{G}) in \e(\ref{t9}), we get
\begin{equation}
\sqrt{R}\dot{R}=-e^{\nu_0(R)}\sqrt{f(r)Re^{2\nu_0}-R+r^3\M}.
\label{collapse}
\end{equation}

\section{Dynamical Evolution of the Collapsing Shells}

The initial density profile, tangential pressure profile and velocity 
profile of the cloud are now as specified above and we have
to evolve the initial data to investigate the possible outcomes of 
the collapse. Initially, all the shells have the scale factor $v(t_i,r)$ as 
unity, with $\dot{v}(t_i,r)<0$, {\it i.e.} an initially collapsing cloud. 
The possible bounce of a shell is given by the change in sign 
of $\dot{v}$. 

\begin{figure}[h!!]
\includegraphics[width=7.0cm,angle=-90]{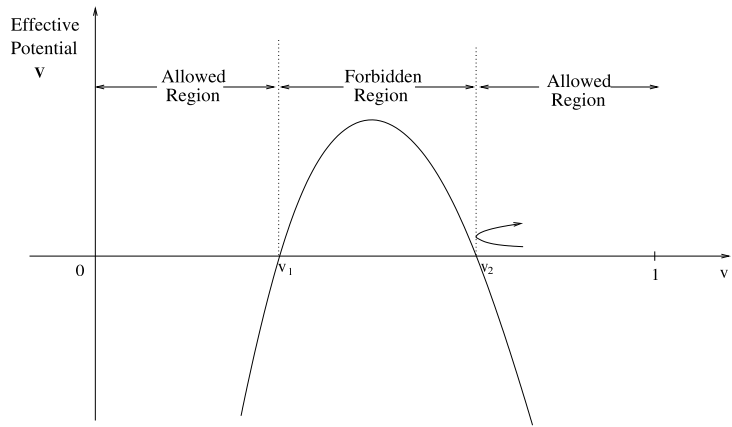}
\caption{Typical effective potential for a bouncing shell.}
\end{figure}

The evolution of a particular shell is deduced from 
\e (\ref{collapse}). Rewriting (\ref{collapse}) in terms of 
$v$, we get
\be
v\dot{v}^{2}= {e^{2\nu_{0}}\,\,[v\,j(r,v)+\M]} \equiv -{V}(r,v),
\label{gendyn}
\eq
where $j(r,v)= [f(r)e^{2\nu_{0}(rv)}-1]/r^{2}$. We call $V(r,v)$ 
as the {\it effective potential} for a shell. A similar effective 
potential has been constructed for studying critical behaviour in scalar 
field collapse in higher dimensions by Frolov 
\cite{frolov}. 
As seen in FIG.1, the allowed regions of 
motion correspond to ${V}(r,v)\leq0$, and dynamics of a shell can be 
studied by finding the turning points. Starting from an initially 
collapsing state ($\dot{v}<0$), there is a rebounce if 
$\dot{v}=0$ (at $v=v_2$ in FIG.1) before the shell has become singular. 
This can happen when ${V}(r,v) = 0$. Hence, to study the various 
possible evolutions for a particular shell, we need to analyze roots 
of the equation ${V}(r,v) = 0$, keeping the value of $r$ fixed. 
We have confirmed numerically that there is no shell-crossing in the 
cloud for a wide range of initial data of interest, which suggests 
that if a particular shell with comoving coordinate $r_a$ bounces then 
all the other shells with $r>r_a$ must also bounce. This implies that 
if central shell at $r=0$ bounces, then the whole cloud 
must also bounce off. This suggests that to investigate the situation 
when the whole cloud is just about to disperse off, it is sufficient 
to study the dispersal of the shells near the center. 
Therefore, to find the threshold of black hole formation and to get 
the scaling relation for the mass of black hole, we need to analyze 
the model only near the center.

With the form of smooth initial data (\ref{mass1}), (\ref{ptheta0}), 
we can integrate (\ref{t7}) on $t=t_i$, close to the center, 
and $\nu_0=\nu_0(R)$ gives
\be
\nu_{0}(R)=\frac{\ptt}{3\Ma}R^2 + \left(\frac{3\ptt+10\Mb}{36\Ma}\right) 
\ptt R^4 +\cdots
\label{nu1}
\eq 
We neglected higher order terms above since we want to consider 
the evolution of shells near $r=0$ only. Near the center, \e(\ref{gendyn})
can be written as
\be
v\dot{v}^{2} \approx A(r,v)
\left[\,P_{5}\,r^2 \,v^5 +\,\left(\frac{2}{3}\frac {\ptt}{\Ma}f(r)\,\right)\,
v^{3} + b(r)\,v + \M \right],
\label{dyn1}
\eq
where $A(r,v)= 1+ (2\ptt r^{2}v^{2})/3m_0$ and 
$P_{5}=({4\ptt^2+\Ma \ptt(3\ptt+10 m_{2}))}/{18\Ma^2}$.
The first factor in $V(r,v)$ i.e function $A(r,v)$, being the 
$|g_{00}|$ term, is always positive and does not contribute to the bounce 
of shells.
 The main features of evolution of cloud derive from the 
second factor in $V(r,v)$, a quintic polynomial having five roots in 
general. Only positive real roots correspond to physical cases. 
Since $V(r,0)=-\mathcal{M}<0$, any region between $R=rv=0$ and the first 
positive zero of $V(r,v)$ always becomes singular during collapse. 
The region between the unique 
positive roots is forbidden as $\dot{v}^{2}<0$ there. For a particular 
shell to bounce it must therefore lie, during initial epoch ($v=1$), in 
a region to the right of the second positive root.

We now study a configuration in which interesting possibilities 
for bounce and collapse arise when $b(r)< 0 \, ; \,\ptt f(r)\,>0$.

If the discriminant of the quintic polynomial is positive, then 
from the Descartes' rule of signs, there are two positive roots 
$\gamma_{1}(r)$ and $\gamma_{2}(r)$ 
\cite{dickson}, 
and the space of allowed dynamics is $[0,\gamma_{1}]$ and 
$[\gamma_{2},\infty)$. The region $(\gamma_{1},\gamma_{2})$ is forbidden. 
Shells in the $[0,\gamma_{1}]$ region initially, {\it always become 
singular}. Shells initially belonging to the region $[\gamma_{2},\infty)$ 
will undergo a bounce and subsequent expansion, starting from initial 
collapse. 
This bounce occurs when their geometric radius approaches 
$R_{bounce}=r\gamma_{2}$. If the initial data is chosen such that 
the discriminant vanishes, then the positive roots are equal and there 
is no forbidden region.

For $b(r)< 0 \, ; \,\ptt f(r)\,>0$ a shell can evolve in three 
different possible ways:\\

i) If the effective potential for a shell has two positive roots,
{\it i.e.} if it crosses the $x$ axis twice in the range $[0,1]$ 
(see FIG.1), then the shell bounces off, as seen by differentiating equation 
(\ref{gendyn}) to get $2v\ddot{v}+\dot{v}^2=-V,_{v}$.
Since near the turning point $V,_{v} < 0$ and $v>0$, we have $\ddot{v}>0$.\\

ii) If $V(r,v)< 0$ in the whole range $[0,1]$, then the shell will 
reach the singularity at $v=0$.\\

iii) When the potential has double roots in $[0,1]$, then it 
indicates that the shell is in critical collapse condition.\\

In order to understand the possible dynamical evolutions clearly, 
we now consider one specific configuration of the initial data, in 
which $\ptt$ and $b(r)$ are kept fixed and $\Ma$ is allowed to vary. 
We take $b(r)=-|b_{0}|$, then close to the center effective potential 
can be written as 
\be
V \approx A\left[P_5r^2 v^5 + \frac {2\ptt}{3\Ma}(1-b_{0}r^2)v^{3}- 
b_{0}v + \Ma (1-\Mb r^2)\right].
\label{dyn2}
\eq

We will see that for the above effective potential, depending 
on the values of initial parameters, three different types of evolutions 
of the collapsing cloud are possible in general.\\

{\bf \it Case A} \\

This type of evolution is depicted in FIG. 2 which shows the 
effective potential for the central shell $(r=0)$ and two other shells. 
Initially we have $v=1$, $\dot{v}<0$ and as the collapse proceeds 
the value of $v$ decreases, reaching a minimum where potential 
becomes zero. Then it again increases so that $\dot{v}> 0$. 
The effective potential moves towards positive side as one goes away 
from the center and the potential has two positive roots for all the 
shells. There is complete bounce of collapsing shells and Minkowski space 
is left behind. \\

\begin{figure}[h!!]
 \includegraphics[width=6.0cm,angle=-90]{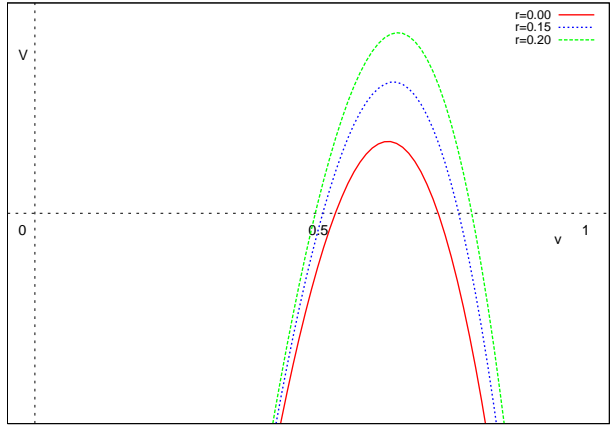}
\caption{Effective potential for different comoving 
radii in {\it Case A} type configuration. Here $m_{0}=1.25$, $\ptt=4.5$, 
$b_{0}=-3.0$ and $m_{2}=0.1.$}
\end{figure}

{\bf \it Case B} \\

As $\Ma$ is increased further, we reach a value at which the potential 
for the central shell just touches the $x$ axis. We call this value as the 
{\it critical value} $m_{0c}$ of the initial parameter $\Ma$ (for the other 
two parameters fixed). This is shown in FIG. 3. The outer shells 
still have positive potential which indicates forbidden region, therefore, all 
those shells will bounce back. This configuration is the boundary point between 
the complete dispersal and black hole formation.\\

\begin{figure}[h!!]
 \includegraphics[width=6.0cm,angle=-90]{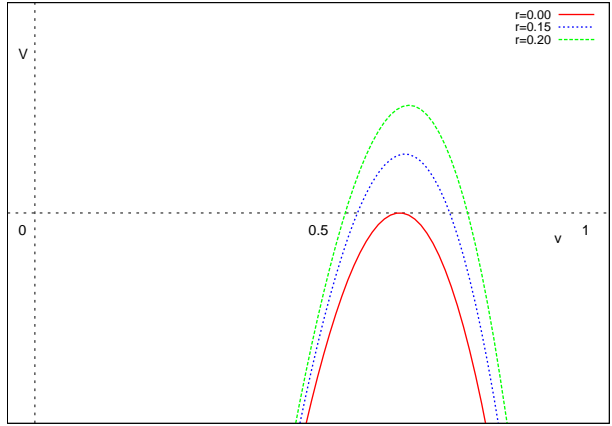}
\caption{Effective potential for different comoving 
radii for {\it Case B} type configuration. Here $m_{0}=1.3333$, $\ptt=4.5$, 
$b_{0}=-3.0$ and $m_{2}=0.1.$ }
\end{figure}

{\bf \it Case C} \\

If we increase $\Ma$ even further, the potential for the central 
shell (FIG. 4) is negative for the whole range of $v$ which will allow the  
central shell to reach the singularity at $v=0$. If we increase $r$, the 
potential minima goes up, and there is a value of $r$ at which the 
effective potential just touches the $x$ axis. We call this radius the 
{\it critical radius $r_c$} of the collapsing cloud for the chosen set of 
initial numbers. Now all the shells from $r=0$ to $r=r_{c}$ will 
reach the singularity and will contribute to the mass of the black hole 
formed, but the shells with comoving coordinate more than $r_c$ 
bounce off. We numerically confirm this by integrating equation 
(\ref{dyn1}). Physical radius for various shells is plotted in FIG. 5. 


\begin{figure}[h!!]
 \includegraphics[width=6.5cm,angle=-90]{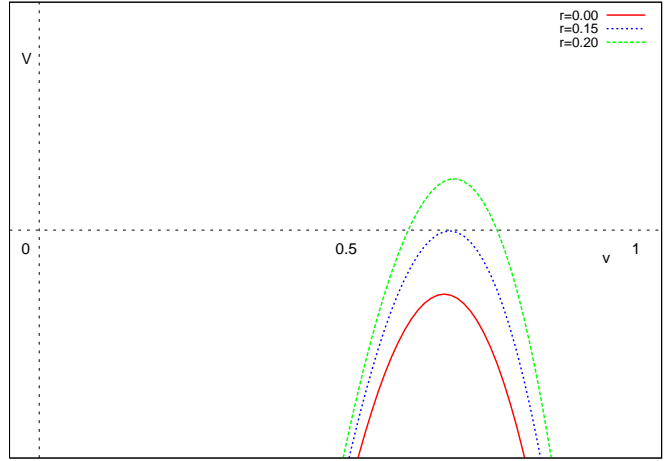}
\caption{Effective potential for different comoving 
radii in {\it Case C} type configuration. Here $m_{0}=1.40$, $\ptt=4.5$, 
$b_{0}=-3.0$ and $m_{2}=0.1.$}
\end{figure}

\begin{figure}[h!!]
\includegraphics[width=6.50cm,angle=-90]{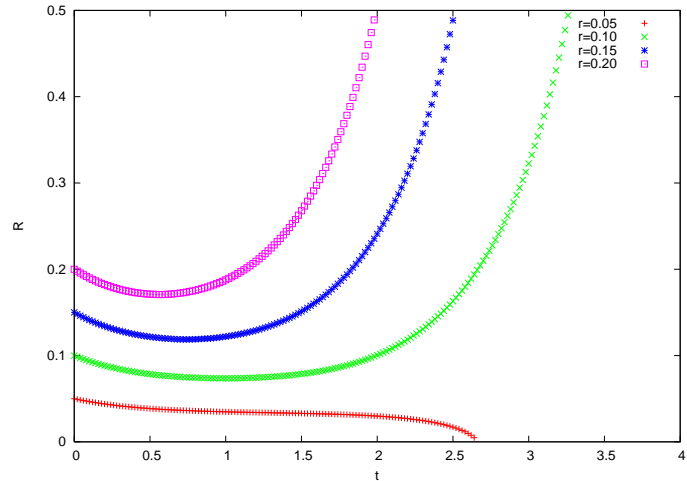}
\begin{small}
\caption{Physical radius of different shells in Case C type evolution,  
where  $\Ma=1.35, |b_{0}|=3.0$, $\Mb=0.1$ and $\ptt=4.5$. 
Shells with comoving radius less than 
the critical radius reach singularity, while those with larger than 
critical value escape to infinity.}
\end{small}
\end{figure}

 We can understand qualitatively why the solution for {\it Case B} is a massless naked singularity. Suppose we are in initial data parameter 
range of {\it Case C} regime and we decrease the value of $m_0$. 
As $m_0$ approaches $m_{0c}$ fewer and fewer shells reach singularity. 
Thus if we approach the threshold of black hole formation from the 
initial data of {\it Case C} region, we can see that when $m_0$ is 
very very close to $m_{0c}$, almost only the central shell reaches 
singularity and other shells are stopped from collapsing and  bounce back 
at the value of $v$ which is the root of the potential for those 
shells. Apparent horizon is given by equation $F=R$, which implies 
$v_{ah}\approx r^2 \,m_0$ for the shells near the center. A shell is 
trapped if it reaches value of $v$ smaller than $v_{ah}$. Now, as 
all outer shells reach a minimum value $v_0(r) >> v_{ah}$, those are 
not trapped. Only the central shell is singular when $m_0$ is 
nearly equal to $m_{0c}$ and trapped surfaces are not formed. 
As $F=r^3 \M(r)$, the mass which contributes towards singularity formation 
is zero which means the singularity is massless and also as the 
trapped surfaces are not formed, the central singularity is visible.

We note that in the collapsing cloud if a shell bounces then 
all the shells with larger value of comoving radius will also bounce, 
{\it i.e.} there is no shell-crossing. To see this, we can integrate 
near the center numerically to obtain $R'$ as well as all the metric 
functions. We see that $R'$ remains positive during the collapse 
evolution which is plotted for one particular initial data set in FIG. 6. 
We find all solutions for various values of initial data and it 
was seen that for the critical solution as well as for solutions  
near the critical point shell cross does not occur.

\begin{figure}[h!!]
\includegraphics[width=6.50cm,angle=-90]{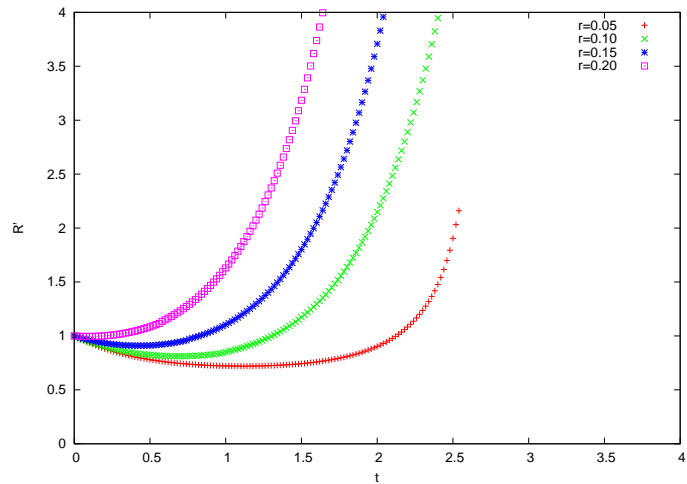}
\begin{small}
\caption{$R'$ for different shells in Case C type evolution,  
where  $\Ma=1.35, |b_{0}|=3.0$, $\Mb=0.1$ and $\ptt=4.5$.}
\end{small}
\end{figure}

The expression for critical radius in terms of the initial 
parameters can be obtained from the condition that at critical radius, 
effective potential just touches the $x$ axis, {\it i.e.}  
the quintic polynomial has double roots. For a quintic polynomial 
$a_{5}x^{5}+a_{3}x^{3}+a_{1}x+a_{0} =0$ , the discriminant is given as 
follows 
\cite{pari},
\begin{figure}[h!!]
\includegraphics[width=6.5cm,angle=-90]{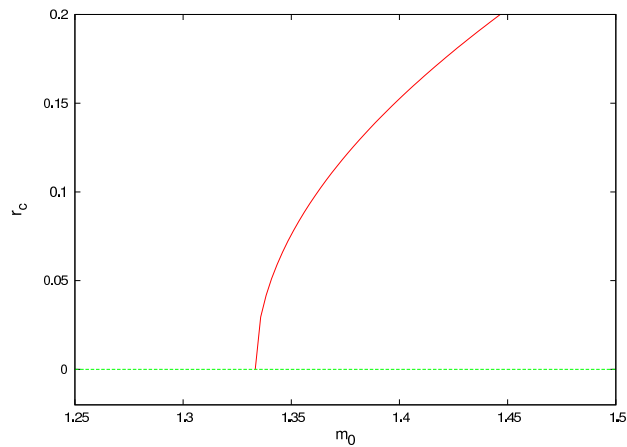}
\caption{Figure represents behavior of {\it critical radius} with 
the density parameter, keeping other parameters values unchanged.}
\end{figure}

\be
\Delta=a_5\left[a_0^2(108a_3^5-900a_1a_3^3a_5+2000a_1^2a_3a_5^2
+3125a_0^2a_5^3)+16a_1^3a_3^4-128a_1^4a_3^2a_5+256a_1^5a_5^2\right].
\eq
The condition that $\Delta$ vanishes at the double root gives, 
from \e (\ref{dyn2}), the expression for critical radius as,

\be
r_{c}^2=\frac{\Ma-m_{0c}}{\Ma (0.833b_{0}+2\Mb)+c_1(0.22b_0^3- 1.04\Ma\ptt )},
\eq

where $c_1=\frac{b_{0}\Ma}{\ptt^2}(3\ptt+10\Mb)$ and 
$m_{0c}=\frac{2b_0^3}{9\ptt}$. Behavior of the critical radius with 
the parameter $\Ma$ can be seen in FIG.7. In the model considered here, 
the mass function, which is twice the Misner-Sharp mass, depends 
only on $r$. The total mass which collapsed to form singularity from 
the regular initial profile is
$\frac{1}{2}r_{c}^3\Ma e^{-m_{1}r_{c}^2}$.
The mass of black hole would then be
\be
M_{BH}= c_{m}(\Ma-m_{0c})^{\frac{3}{2}},
\eq
where $c_m$ is a constant near the threshold. We can fix any of 
the two parameters  $\Ma$, $b_{0}$, and $\ptt$ and vary the third one 
to obtain the expression for critical radius. Following the same 
procedure as earlier, it is easily seen that for $b_{0}$ and $\ptt$ 
the same scaling relation and exponent exists near the threshold of 
the black hole formation. 
Therefore, in general, for a parameter $\xi$, we can write  
$M_{BH}= c_{\xi}|\xi-\xi_{c}|^{\frac{3}{2}}$.

\section{Discussion and Conclusions}

We like to point out the similarities in the behaviour 
of this model with the critical behavior in scalar field gravitational 
collapse 
\cite{choptuik}, 
and other matter 
models 
\cite{radiation}:

i) The three types of evolutions discussed in {\it Case A-C} are 
very much analogous to the {\it subcritical}, {\it Critical} and 
{\it super-critical} evolutions studied in the above models. In case A the 
whole cloud disperse off and in Case C black hole forms whose mass can 
be controlled by an initial parameter. Case B is the boundary between 
the case A and Case C configuration which is like the critical 
solution in the above cases.

ii) A power law behaviour is seen for the mass of the black hole 
near the threshold of its formation. Infinitesimal mass black holes 
can be formed in this model if the initial parameters are finely tuned.

iii) The same value of exponent is observed for different initial 
parameters which is like universality of the critical solution.\\

Our purpose here is to investigate how the initial parameters 
determine the evolution and the end state of collapse. The metric 
function $\nu$, in spite being a restrictive choice, makes the model 
tractable. The other class of tangential pressure models, {\it i.e} 
non-static Einstein cluster has a different form of the function 
$\nu$ in comoving coordinates and it can not be written as function 
of $R$ only. We have observed similar behavior at threshold of 
black hole formation in Einstein cluster as well and the same value 
of the exponent was seen 
\cite{mahajan}.
The effective potential method for calculating the critical 
exponent applies to this case as well as to non-static Einstein cluster 
\cite{mahajan}. 
This suggests that the method applies at least for all 
mass-conserving systems. However, the system considered here has a 
limitation, and is different from the models considered so far 
for critical behaviour in one respect that it has no radial stress 
which enables shells to interact directly with each other. 
It will be interesting to explore whether the 
method we used here could be applied for studying such behaviour 
in the models with a non-vanishing radial pressure as well.\\


{\bf Acknowledgments}

S. Gutti, R. Goswami and A. Madhav are thanked for 
useful discussions.

\section*{APPENDIX A : BLACK HOLE FORMATION FOR COLLAPSE IN CASE C CONFIGURATION}

We can rewrite \e(\ref{collapse}) as 
\begin{equation}
\sqrt{v}\dot{v}=-\sqrt{e^{4\nu_0}vb_0+e^{2\nu_0}\left(v^3h(rv)+\M\right)},
\label{collapse1}
\end{equation}
where
\begin{equation}
h(R)=\frac{e^{2\nu_0(R)}-1}{R^2}.
\label{h}
\end{equation}
Integrating the above equation, we get
\begin{equation}
t(v,r)=\int_v^1\frac{\sqrt{v}dv}{\sqrt{e^{4\nu_0}vb_0+e^{2\nu_0}
\left(v^3h(rv)+\M\right)}}
\label{eq:scurve1}
\end{equation}
The time of singularity for a shell at a comoving coordinate 
radius $r$ is the time when the 
physical radius $R(r,t)$ becomes zero. 
The shells collapse consecutively, that is one after the other to the 
center as there are no 
shell-crossings.
Taylor expanding the above function around $r=0$, we get 
\be 
t(v,r)=t(v,0)\;+\left.r\;\frac{d t(v,r)}{dr}\right|_{r=0}  
+\left.\frac{r^2}{2!}\;\frac{d^{2}t(v,r)}{d^2{r}^2}\right|_{r=0}
\label{eq:scurve2}
\eq
Let us denote 
\be
\X_{n}(v)=\left.\frac{{d} ^{n} t(v,r)}{{d} r^{n}}\right|_{r=0}  
\eq
As we have taken the initial data with only even powers of $r$, 
the first derivatives
of the functions appearing in above equations vanish at $r=0$, hence
we have
\be
\X_{1}(v)=0
\eq
Now we can express the next coefficient $\X_{2}$ as
\be
\X_{2}(v)=-\int_{v}^{1}\frac{\sqrt{v}dv[\beta v^5 + 2 \Ma p_{\theta_{2}}
v^2+ 9 \Ma^2(b_{02}v+\Mb)]}{9m_0^2\,(hv^3+b_{00}v+ M)^{\frac{3}{2}}}
\label{eq:scurve5}
\eq
where
\be
\beta=6p_{\theta_{2}}^2+24 \Ma b_{02}p_{\theta_{2}}-3 \Ma \rho_{2}p_{\theta_{2}}\eq

We need to determine now whether it is possible to have
families of future directed outgoing null geodesics coming out
of the singularity. In the case when
such families do exist which terminate in the past at the singularity,
and which could reach outside observers, then the singularity will
be visible. In the case otherwise it is hidden within the
black hole. Another way to look at this is through the apparent horizon
and formation of trapped surfaces in the spacetime. As the
collapse evolves, if the trapped surfaces form well in advance to 
the formation of the singularity, then the same will be covered.
On the other hand, if the trapped surface formation is sufficiently 
delayed during the collapse then the singularity may be naked.
The apparent horizon within the collapsing cloud is given by the \e,
$R/F=1$, which gives the boundary of the trapped surface region 
of the space-time. If the neighborhood of the center gets 
trapped earlier than the singularity, then it is covered, 
otherwise it is naked with non-spacelike future directed 
trajectories escaping from it.

In order to consider the possibility of existence 
of such families, and to examine the nature of the singularity 
occurring at $R=0, r=0$ in this model, let us consider the 
outgoing null geodesic equation which is given by
\begin{equation}
\frac{dt}{dr}=e^{\psi-\nu}
\label{eq:null1}
\end{equation}
We now use here a method 
which is 
similar to that given in
\cite{dim}.
The singularity curve is given by $v(t_s(r),r)=0$, which corresponds 
to $R(t_s(r),r)=0$. Therefore, if we have any future directed 
outgoing null 
geodesics terminating in the past at the singularity, we must have 
$R\rightarrow 0$ as $t\rightarrow t_s$ along the same. 
Now writing \e 
(\ref{eq:null1}) 
in terms of variables $(u=r^\alpha,R)$, we have
\begin{equation}
\frac{dR}{du}=\frac{1}{\alpha}r^{-(\alpha-1)}R'\left[1+\frac{\dot{R}}{R'}
e^{\psi-\nu}\right]
\label{eq:null2}
\end{equation}

Now in order to get tangent to the null geodesic in the $(R,u)$ plane,
we choose a particular value of $\alpha$ such that the geodesic equation 
is expressed only in terms of $\left(\frac{R}{u}\right)$. 
A specific value of alpha is to be chosen which enables us to calculate
the proper limits at the central singularity.
For example, 
for $\X_1(0)\ne0$ case,
we can choose $\alpha=\frac{5}{3}$ and using \e (\ref{t9}), 
(and considering that $\dot{R}<0$), we get
\begin{equation}
\frac{dR}{du}=\frac{3}{5}\left(\frac{R}{u}+\frac{\sqrt{\M_0}\X_1(0)}
{\sqrt{\frac{R}{u}}}\right)\left(\frac{1-\frac{F}{R}}{\sqrt{G}[\sqrt{G}
+\sqrt{H}]}\right)
\label{eq:null3}
\end{equation}
In the tangential pressure collapse model discussed in the 
previous section we have $\X_1(0)=0$, and hence
we choose $\alpha=\frac{7}{3}$ so that when in limit $r\rightarrow 0, 
t\rightarrow t_{s}$ we get the value of tangent to null geodesic in 
the $(R,u)$ plane
\be
\frac{dR}{du}=\frac{3}{7}\left(\frac{R}{u}+\frac{\sqrt{M_{0}}\X_{2}(0)}
{\sqrt{\frac{R}{u}}}\right)\frac{(1-\frac{F}{R})}{\sqrt{G}(\sqrt{G} 
+\sqrt{H})}
\label{eq:null4}
\eq

Now note that for any point with $r>0$ on the singularity curve 
$t_s(r)$, we have $R\to 0$ whereas $F$ (interpreted as mass of the
object within the comoving radius $r$) tends to a finite positive
value once the energy conditions are satisfied. Under the situation,
the term $F/R$ diverges in the above equation, and all such points 
on the singularity curve will be covered as there will be no outgoing
null geodesics from such points.

Hence we need to examine the central singularity at $r=0,R=0$
to determine if it is visible or not. That is, we need to determine
if there are any solutions existing to the outgoing null geodesics 
equation, which terminate in the past at the singularity and in future
go to a faraway observer, and 
if so under what conditions these exist. Note 
that if any outgoing null geodesics terminate at the singularity 
in the past then along the 
same, in the limit as $r\rightarrow 0, t\rightarrow t_{s}$ 
we then have from equation (26) $\dot{R} =0$, therefore $H=0$ and  $G=1$ 
in this limit as $F/R$ vanishes.
Let now $x_{0}$ be the tangent to the null geodesics in $(R,u)$ plane, 
at the central singularity, then it is given by,
\begin{equation}
x_0=\lim_{t\rightarrow t_s}\lim_{r\rightarrow 0} 
\frac{R}{u}=\left.\frac{dR}{du}\right|_{t\rightarrow t_s;r\rightarrow 0}
\end{equation}
Using \e (\ref{eq:null4}), we get,
\begin{equation}
x_0^{\frac{3}{2}}=\frac{7}{4}\sqrt{\M_0}\X_2(0)
\end{equation}
In the $(R,u)$ plane, the null geodesic equation will be,
\be
R=x_0u
\eq

It follows that if $\X_2(0)>0$, then that implies that $x_0>0$, 
and we then have radially  outgoing null geodesics coming out from the 
singularity, making the central singularity to be a visible one. 
On the other hand, if 
$\X_2(0)\le0$, we will have a black hole solution. 
Now for the initial data in the range of Case C, we calculate 
$\X_2(0)$ and find that it remains negative which implies that a black hole 
is necessarily formed.

\end{document}